\documentclass[aps,prd,preprintnumbers]{revtex4}
\usepackage{amsfonts}
\usepackage{hyperref}
\usepackage{amsmath}
\usepackage{amssymb}
\usepackage[english]{babel}
\usepackage{graphicx}
\usepackage{epsfig}
\usepackage{bm}
\usepackage{longtable}
\usepackage{verbatim}
\usepackage{longtable}

\newcommand{\mathsym}[1]{{}}

\begin{document}

\newcommand{\nn}{\nonumber}
\newcommand{\beqa}{\begin{eqnarray}}
\newcommand{\eeqa}{\end{eqnarray}}
\newcommand{\fs}{\displaystyle{\not}} 
\newcommand{\ev}[1]{\left<#1\right>}
\newcommand{\bk}[2]{\left<#1|#2\right>}
\newcommand{\bok}[3]{\left<#1|#2|#3\right>}
\newcommand{\mev}{\mbox{ MeV}}
\newcommand{\red}[1]{\textcolor{red}{#1}}
\newcommand{\blue}[1]{\textcolor{blue}{#1}}
%
\def\al{\alpha}
\def\be{\beta}
\def\ga{\gamma}
\def\de{\delta}
\def\ep{\epsilon}
\def\Ga{\Gamma}
\def\Up{\Upsilon}
\def\Dot{\!\cdot\!}
\def\pr{\prime}

\title{$\chi_b(3P)$ splitting predictions in potential models}
\author{Claudio O. Dib${}^{}$}
\email{claudio.dib@usm.cl}
\author{Nicol\'as A. Neill${}^{}$}
\email{nicolas.neill@gmail.com}
\affiliation{${}^{}$Universidad T\'ecnica Federico Santa Mar\'{\i}a\\
and\\
Centro Cient\'{\i}fico-Tecnol\'ogico de Valpara\'{\i}so\\
Casilla 110-V, Valpara\'{\i}so, Chile}
\date{\today }

\begin{abstract}
Recently a new state of the $b\bar{b}$ system has been observed by the ATLAS Collaboration, with a mass of $10530 \pm 5 \pm 9\mbox{ MeV}$.
This state has been identified with the $n=3$ P-wave radial excitations of the $b\bar b$ system with parallel quark spins, called $\chi_b(3P)$. The measured value of the mass corresponds to the average of the $J=0, 1$ and $2$ states, while the splittings of these states are not yet resolved.
In this work we present predictions from different potentials models for the values of these splittings.
\end{abstract}

\maketitle

\section{Introduction}\label{sec:intro}

Heavy quarkonia, or mesons formed as heavy quark-antiquark bound states, constitute a valuable ground to study strong interactions, because they are related to both the perturbative and non-perturbative regimes of QCD.  In heavy quarkonia energies are marginally high enough for perturbative methods to become useful, while the confining, non perturbative aspect of the interaction, is also of crucial importance.

Recently a new state of the $b\bar{b}$ system has been observed by the ATLAS Collaboration, with a spin-weighted average (``barycenter") mass of $10530 \pm 5 \pm 9\mbox{ MeV}$ \cite{Aad:2011ih}.
This state has been identified with the $n=3$ radial excitation of the $b\bar{b}$ system with angular momentum numbers $s=1$, $\ell=1$ (hence $J=0,1,2$), called $\chi_b(3P_J)$.
The three states $J=0,1,2$ are closely spaced in mass, so the measured value of the mass corresponds to their average, while the splittings of these states are not yet resolved.
In this work we present predictions from different potentials models for the values of these splittings.

Since the discovery of the $J/\psi$ meson in 1974, understood as a $(c\bar c)$ system, many potential models have been proposed to describe the interaction between the quark and antiquark in a bound state \cite{
Eichten:1974af,Pumplin:1975cr,Quigg:1977dd,Eichten:1978tg,Richardson:1978bt,Eichten:1979ms,Buchmuller:1980su,Martin:1980rm,Buchmuller:1981aj,Gupta:1981pd,
Gupta:1982kp,Gupta:1982qc,Gupta:1984um,Gupta:1986xt,Born:1989iv,Gupta:1993pd,Ding:1995he,Ebert:2002pp,Barnes:2003vb,Gonzalez:2003gx,
Eichten:2004uh,Barnes:2005pb,Eichten:2005ga,Lakhina:2006vg}.
One of the first proposed models was the so called \emph{Cornell potential} \cite{Eichten:1974af,Eichten:1978tg,Eichten:1979ms}, which is a Coulomb-plus-linear term combination that takes into account general properties expected from the interquark interaction: a Coulombic behavior at short distances and a linear confining term at long distances, representing the perturbative one-gluon exchange and the non-perturbative chromoelectric flux tube of confinement, respectively.
Other models based on phenomenological grounds include a logarithmic potential \cite{Quigg:1977dd} and a non-integer power law potential \cite{Martin:1980rm}.
Subsequently, elements from QCD were included in different ways in the potential formulation, from the inclusion of the running of the QCD coupling constant in the Coulombic interaction \cite{Richardson:1978bt}, to a derivation of the short distance quark-antiquark potential from perturbative QCD, where spin-dependent interactions naturally appear \cite{Buchmuller:1980su,Gupta:1981pd,Buchmuller:1981aj,Gupta:1982qc,Gupta:1982kp,Gupta:1984um,Gupta:1986xt}.
A modification of the Cornell potential to take into account a saturation effect in the linearly growing confining part has also been proposed, inspired in lattice data \cite{Born:1989iv,Gonzalez:2003gx,Ding:1995he}.
In the context of these potential models, a successful description of many properties of quarkonia has been done, both for mass spectrum and decay widths.
A comprehensive review of the status of heavy quarkonium can be found in Ref. \cite{Brambilla:2010cs}. For specific studies of P-wave splittings in $b\bar b$, see also  Ref.~\cite{chisplit}.

Even though the $b$ quarks are heavy compared to the scale of confinement in QCD, the $b$-$\bar b$ interaction is sufficiently strong to make the bound state somewhat relativistic. Quark model calculations give typical velocities $v_q\sim 0.25-0.4$.
Consequently, one expects relativistic effects to be important, and at a fundamental level the very use of a potential, which represents an instantaneous static interaction, is questionable. Nevertheless, potential models have been successful in predicting mass spectra and decays in quarkonia, and are particularly useful in describing the mass splitting among different states of orbital angular momentum and spin. 

Here we study the prediction on the mass splitting in $\chi_b(3P)$ from a set of potential models that have been able to predict other quarkonium spectra with relative success. 
We compare the average masses of known parts of the spectra in P-wave states with the results given by each of the potential models.
An important issue to consider in the model is the separation between scalar vs. vector nature of the quark-antiquark interaction, as the spin-dependent interaction is sensitive to this separation. Since this separation is purely phenomenological, we must first use as input a known part of the spectrum in order to fit the parameters that model this separation;  we then obtain results for other parts of the known spectrum, which we use to check the validity of the model. Then, taking into consideration the degree of success of these results, we adventure in predicting the yet unknown mass splitting in the $\chi_b(3P)$ state within each of these models.

In Section \ref{models} we briefly describe the potential models that we consider in our analysis.
In Section \ref{results} we present the results obtained from the different models including the predictions for the splittings of the $\chi_b(3P)$ state.
Finally in Section \ref{conclusions} we state our conclusions.

\section{Potential models}\label{models}

Rigorously speaking, a potential in a bound state is a static instantaneous interaction. Such a description is good only in the non relativistic limit. A systematic relativistic expansion of the interaction provides additional terms that are corrections that depend on the orbital angular momentum and spin of the bound particles. The expansion should represent charmonium and bottomonium rather well since the motion of their constituents is not highly relativistic. In what follows we consider four representative models given in terms of a spin-independent, static central potential, where relativistic corrections are incorporated as the aforementioned expansion. We call these models a) the Cornell model b) the Screened Cornell model c) the Richardson model, and d) the p-QCD (perturbative QCD) model, which we describe in what follows. In addition, for the first two models we consider two variations, henceforth called Cornell-I and Cornell-II, and Screened Cornell-I and Screened Cornell-II, respectively, depending on the separation of the potential into a vector exchange and a scalar exchange, as explained below. 

\subsection{The Cornell model}

The \emph{Cornell potential} \cite{Eichten:1974af} was originally proposed to describe masses and decay widths of charmonium states.
It takes into account the following features of the interaction between the constituents: (i) a Coulomb-type interaction which should describe the short distance regime mediated by a single gluon exchange in perturbation theory, and (ii) a quark confining interaction that dominates at long distances. 
Explicitly, the potential is given by
\begin{eqnarray} \label{e:cornell-a}
V(r) = -\frac{\beta}{r} + \frac{r}{a^2} + V_0,
\end{eqnarray}
where the first term corresponds to the Coulombic interaction and the second is the interaction responsible for confinement that arises from the color field flux tube between the quarks.
The coefficients $a$ and $\beta$ are adjusted to fit the charmonium spectrum, but with the assumption that, roughly speaking, it should be valid for all other heavy quarkonia. As such, the flavor dependence should arise solely from the mass of the bound quarks.

Many fits have been done to adjust these parameters \cite{Eichten:1974af,Quigg:1979vr,Eichten:1979ms}.
In this work we will consider the most recent values: $\beta=0.52$, $a=2.34\mbox{ GeV}^{-1}$, $m_c=1.84\mbox{ GeV}$ and $m_b = 5.17\mbox{ GeV}$ \cite{Eichten:1979ms}, where $\beta$, $a$ and $m_c$ were fitted using as inputs the mass differences of the charmonium states $J/\psi(1S)$, $\psi(2S)$ and the $\chi_c(1P)$ spin-weighted barycenter, as well as the requirement that the charm quark mass should be close to the lightest charmed-meson mass. The value of $m_b$ was chosen taking into account limits based on the $\Upsilon$ masses.

The potential in Eq.\ (\ref{e:cornell-a}) represents the static limit of the interaction. A relativistic expansion will include additional terms, that will constitute smaller corrections in the case of heavy quarkonia, and some of them  will depend on the orbital angular momentum and spin of the constituents.
These terms are essential in order to reproduce or predict the mass splittings of quarkonium states that differ in angular momentum, such as the 
$\chi_b(3P_J)$, for $J=0, 1, 2$.

A general expression for the spin-dependent part of the potential of a $q\bar q$ bound state in QCD using a Wilson loop approach was calculated to order $1/m^2$: \cite{Eichten:1980mw, Pineda:2000sz}
\begin{eqnarray} \label{e:vsd1}
V_{SD}(r) & = & \left[\frac{\mathbf{S_1}\cdot \mathbf{L_1}}{2m_1^2} - \frac{\mathbf{S_2}\cdot \mathbf{L_2}}{2m_2^2} \right]
            \left[\frac{dV(r)}{rdr}+2\frac{dV_1(r)}{rdr}\right] + \frac{(\mathbf{S_2}\cdot\mathbf{L_1}-\mathbf{S_1}\cdot\mathbf{L_2})}{2 m_1 m_2} \frac{dV_2(r)}{rdr}\nonumber\\
    & &        + \frac{1}{6 m_1 m_2}(6\mathbf{S_1}\cdot \mathbf{\hat{r}} \mathbf{S_2}\cdot \mathbf{\hat{r}} -2\mathbf{S_1}\cdot\mathbf{S_2} )V_3(r) + \frac{2}{3 m_1 m_2} \mathbf{S_1}\cdot\mathbf{S_2} \nabla^2 V_2(r),
\end{eqnarray}
where $V(r)$ is the spin-independent part of the potential, while $V_1(r)$, $V_2(r)$ and $V_3(r)$ are the spin-dependent parts, that can be expressed as expectation values of color electric and magnetic fields. These potentials are not all independent, as pointed out by Gromes \cite{Gromes:1984ma}.
Lorentz invariance imposes the relation  $V(r)+V_1(r)-V_2(r)= {\cal C}$, where $\cal C$ is an irrelevant integration constant. We can thus eliminate $dV_1(r)/dr$ from Eq.\ (\ref{e:vsd1}) and, in the 
equal-mass case, $m_1=m_2\equiv m$, we can rewrite it as:
\begin{equation} \label{e:vsd2}
V_{SD}(r)  =  \frac{\mathbf{S}\cdot \mathbf{L}}{2m^2}
            \left[-\frac{dV(r)}{rdr}+4\frac{dV_2(r)}{rdr}\right] + \frac{1}{12 m^2}(6\mathbf{S}\cdot \mathbf{\hat{r}} \mathbf{S}\cdot \mathbf{\hat{r}} -2\mathbf{S}\cdot\mathbf{S} )V_3(r) + \frac{1}{6 m^2} \left(2\mathbf{S}\cdot\mathbf{S}-3\right) \nabla^2 V_2(r),
\end{equation}
where $\mathbf{S}\equiv \mathbf{S_1}+\mathbf{S_2}$ and $\mathbf{L}\equiv \mathbf{L_1}=-\mathbf{L_2}$. The terms in Eq.\ (\ref{e:vsd2}) are referred to as spin-orbit, tensor and hyperfine interaction, respectively. 
Notice that only the spin-orbit and tensor interactions are responsible for the mass splittings of the $\chi_b(nP_J)$.

Now, the otherwise independent potentials $V(r)$, $V_2(r)$ and $V_3(r)$ are further related if one assumes that the quark-antiquark potential due to the strong interaction arises from the effective exchange of a scalar and a vector particle. This assumption is clearly part of the modeling, as there could be further effective exchanges, but if we stick to it, the quark-antiquark lagrangian has the form:
\begin{eqnarray}
L_{int} = \tilde{s}(q^2)\bar{u}u\bar{v}v + \tilde{v}(q^2)\bar{u}\gamma_\mu u\bar{v}\gamma^\mu v. 
\end{eqnarray}
The expansion of this expression in powers of $v^2/c^2$ gives a Breit-Fermi form of the spin-dependent potential \cite{Schnitzer:1975ux,Dib:1987cj}:
\begin{eqnarray} \label{e:vsd3}
V^{(eff)}_{SD}(r) & =& \frac{\mathbf{S}\cdot \mathbf{L}}{2m^2}\left[ -\frac{dv(r)+ds(r)}{rdr} + 4\frac{dv(r)}{rdr} \right]
            + \frac{1}{12m^2}(6\mathbf{S}\cdot \mathbf{\hat{r}} \mathbf{S}\cdot \mathbf{\hat{r}} -2\mathbf{S}\cdot\mathbf{S} )
            \left[ \frac{dv(r)}{rdr} - \frac{d^2v(r)}{dr^2}\right]\nn\\
          &  &  + \frac{1}{6m^2}(2\mathbf{S}\cdot \mathbf{S}-3)\nabla^2v(r).
\end{eqnarray}
Comparing \ Eqs.\ (\ref{e:vsd2}) and (\ref{e:vsd3}), one can make the following identifications, which relate $V(r)$, $V_2(r)$ and $V_3(r)$:
\begin{eqnarray}
V(r) = v(r)+s(r),\ \ \ \ \ V_2(r)=v(r),\ \ \ \ \ V_3(r) = \frac{dv(r)}{rdr} - \frac{d^2v(r)}{dr^2}.\nonumber
\end{eqnarray}
We then see that the spin-independent potential $V(r)$ is composed by the vector and scalar parts, while the spin-dependent parts, $V_2(r)$ and $V_3(r)$, are given in terms of the vector part only.

To leading order in the wave function obtained from the spin-independent potential $V(r)$ (e.g.  Eq.\ \ref{e:cornell-a}), one can express the spectrum of the $\chi_b(nP_J)$, for a fixed radial excitation $n$,  as: 
\begin{eqnarray} \label{e:splittings}
M(P_2) & = & \bar{M} + a - 2b/5,\nonumber\\
M(P_1) & = & \bar{M} - a + 2b,\\
M(P_0) & = & \bar{M} - 2a - 4b,\nonumber
\end{eqnarray}
where $a$ and $b$ are the expectation values of the radial functions in the spin-orbit and tensor terms, respectively:
\begin{eqnarray}
a & = & \frac{1}{2m^2} \left<-\frac{ds}{rdr} + 3 \frac{dv}{rdr}\right>,\nonumber\\
b & = & \frac{1}{12m^2} \left< \frac{dv}{rdr} - \frac{d^2v}{dr^2}\right>.\nonumber
\end{eqnarray}
As such, the potential $V(r)$ provides an explicit form for the sum $v(r)+s(r)$, and for the bound state wave function to leading order.
However, to obtain the explicit form of the spin-dependent potential, Eq.\ (\ref{e:vsd3}), it is necessary 
to know, or to define within the model, the separation of $V(r)$  into the scalar and vector parts  $s(r)$ and $v(r)$.

For the Cornell potential, Eq.\ (\ref{e:cornell-a}), one can argue that a reasonable  separation is to assign the Coulombic term to the vector part, coming from a single gluon exchange at short distances, and the linear term to the scalar part, coming from the flux tube at long distances:
\begin{eqnarray} \label{e:sepcor1}
v(r) = -\frac{\beta}{r},\ \ \ \ \ s(r) = \frac{r}{a^2}.
\end{eqnarray}
Here we call this prescription the \emph{Cornell-I} model. 

Alternatively one can relax this assumption by introducing two phenomenological parameters, $\eta_s$ and $\eta_v$, that define the combinations of Coulomb and linear terms of the scalar and vector potentials (where the case $\eta_s=\eta_v=1$ corresponds to Eq.\ \ref{e:sepcor1}):
\begin{eqnarray} \label{e:cornell-b}
s(r) & = & -(1-\eta_v) \frac{\beta}{r} + \eta_s\frac{r}{a^2},\nonumber\\
v(r) & = & -\eta_v \frac{\beta}{r} + (1-\eta_s)\frac{r}{a^2}.
\end{eqnarray}
One can then find the values of $\eta_s$ and $\eta_v$ that provide the best fit to a part of the spectrum. In this study we look for the values that minimize the following $\chi^2$ function, using the mass splittings in the experimentally known $\chi_b(1P_J)$ and $\chi_b(2P_J)$: 
\beqa \label{e:chi2}
\chi^2 = \sum_{\substack{n=1,2 \\ij=21,10}} \left(\frac{\Delta M_{ij} (nP)^{(\mbox{model})} - \Delta M_{ij}(nP)^{(\mbox{exp})}}{\sigma_{ij}(nP)^{(\mbox{exp})}}\right)^2,
\eeqa
where 
\begin{equation}\label{SplitD}
\Delta M_{21}{(1P)} \equiv M(1P_2) - M(1P_1), \quad \Delta M_{10}{(1P)}  \equiv M(1P_1) - M(1P_0)
\end{equation}
are the mass splittings in the $\chi_b(1P_J)$,  and similarly are the $\Delta M_{ij} (2P)$ defined for the $\chi_b(2P_J)$. 
Following this procedure, we find the optimal values $\eta_s =1$ and $\eta_v = 0.923$. We call this prescription the \emph{Cornell-II} model.

It is interesting to notice that the optimal values of $\eta_s$ and $\eta_v$ are remarkably close to the Cornell-I prescription, which is the one commonly assumed in the literature.

\subsection{The Screened Cornell model}

A variation of the Cornell potential, which we call the \emph{Screened Cornell potential},  has been used to include the effect of saturation of the strong interaction at long distances that appears in lattice data \cite{Born:1989iv}:
\begin{eqnarray} \label{e:screened1a}
V(r) = \left(\frac{-\beta}{r}+\frac{r}{a^2}\right)\left(\frac{1-e^{-\mu r}}{\mu r}\right),
\end{eqnarray}
where $\mu$ is the \emph{screening parameter}.
This potential behaves like a Coulomb potential at short distances but, unlike in the previous model, it tends to a constant value for large $r$ (namely, for $r\gg \mu^{-1})$.
In other words, the linearly growing confining potential flattens to a finite value at large distances, corresponding to the  saturation of $\alpha_S$ to a finite value for decreasing $Q^2$ \cite{Cornwall:1989gv,Papavassiliou:1991hx,Aguilar:2001zy}. This effect should be due to the creation of virtual light quark pairs that screen the interaction between the bound quarks at long distances.
The values of $\beta$, $a$ and $\mu$ we use here are those of Ref.\ \cite{Gonzalez:2003gx} and are shown in Table \ref{t:parameters1}.
Of these values, $\beta$ and $\mu$ are intrinsic to the model, while $a$, $m_c$ and $m_b$ were fixed by the authors in order to reproduce the $J/\psi$ mass in $c\bar c$ and the $\Upsilon(1S)$ and $\Upsilon(2S)$ in $b \bar b$.

For the potential in Eq.\ (\ref{e:screened1a}), we can separate the scalar and vector parts in the same common way as in the Cornell potential (that is, the Coulombic term as vector and the linear term as scalar), a prescription we call here the \emph{Screened Cornell-I} model. 

Alternatively, just as before, we can introduce two phenomenological separation parameters, $\eta_s$ and $\eta_v$, which are fitted by the same optimization procedure:
\begin{eqnarray} \label{e:screened1b}
s(r) & = & \left(-(1-\eta_v) \frac{\beta}{r} + \eta_s\frac{r}{a^2}\right) \left(\frac{1-e^{-\mu r}}{\mu r}\right),\nonumber\\
v(r) & = & \left(-\eta_v \frac{\beta}{r} + (1-\eta_s)\frac{r}{a^2}\right) \left(\frac{1-e^{-\mu r}}{\mu r}\right).
\end{eqnarray}
The optimal values we find in this prescription are $\eta_s = 0.810$ and $\eta_v = 1$. We call this prescription the \emph{Screened Cornell-II} model.

\subsection{The Richardson model}

The \emph{ Richardson potential} is another well known model that incorporates the features of asymptotic freedom at short distances and linear confinement at long distances \cite{Richardson:1978bt}.
With a minimal interpolation between these two asymptotic behaviors, the Richardson potential is obtained:
\begin{eqnarray} \label{e:richardson}
V(r) = \frac{8\pi}{33-2n_f} \Lambda \left[\Lambda r - \frac{f(\Lambda r)}{\Lambda r} \right],
\end{eqnarray}
with
\begin{eqnarray}
f(t) = 1-4\int_1^\infty \frac{dq}{q} \frac{e^{-qt}}{\ln^2(q^2-1)+\pi^2}.\nn
\end{eqnarray}
Here $n_f$ is the number of light quarks relevant to the renormalization scale, taken equal to three, while $\Lambda=0.398\mbox{ GeV}$ is the scale of interpolation between the two asymptotic regimes and $m_c = 1.49\mbox{ GeV}$ is the constituent mass of the charm quark in the model. These values have been fitted to reproduce the mass of the charmonium states $J/\psi(1S)$ and $\psi(2S)$. For bottomonium the same value of $n_f$ and $\Lambda$ should be used, setting the bottom quark mass at $m_b=4.8877\mbox{ GeV}$ in order to reproduce the current $\Upsilon(1S)$ mass.
According to Richardson, the value of $n_f$ is kept equal to three (i.e. the number of light quarks), since the Appelquist Carazzone theorem \cite{Appelquist:1974tg} implies that the effect of quarks heavier than the energy scale that determines the dynamics should be small (the latter is related to the binding energy or the inverse of the radius). With these values for $\Lambda$, $n_f$ and $m_b$, the $\Upsilon(2S)$ mass is predicted with good agreement with experiment.

In contrast to the Cornell potential, the Richardson potential has no obvious separation into scalar and vector parts. In order to be able to determine the splittings with this potential, one needs this separation. The following phenomenological separation has been used in previous works \cite{Dib:1987cj}:
\begin{eqnarray} \label{e:richardsonsep}
v(r) & = & V(r) e^{-r^2/a_r^2},\nonumber\\
s(r) & = & V(r) (1-e^{-r^2/a_r^2}).
\end{eqnarray}
Here $V(r)$ is the Richardson potential of Eq.\ (\ref{e:richardson}) and $a_r$ is a phenomenological length scale that separates  the vector character at short distances and scalar character at large distances. Following the same $\chi^2$ optimization procedure as before, the value $a_r = 0.1782\mbox{ fm}$ is found.

\begin{table}[t]
\caption{Best values of the parameters for each of the potential models, according to previous  authors, except $m_b$ in the Richardson model, which was updated with the current experimental data, and the separation parameters $\eta_s$, $\eta_v$ and $a_r$ obtained in our own fits.} \label{t:parameters1}
\centering
\begin{tabular}{l c c c c c c c c c c }
\hline \hline
                                  & Equation    \ \ \          & $\beta$    &  $a\mbox{ GeV}^{-1}$ &  $m_b\mbox{ GeV}$  &  $V_0\mbox{ GeV}$ &  $\mu\mbox{ fm}^{-1}$ &   $\eta_s$   & $\eta_v$ &    $\Lambda\mbox{ GeV}$  &  $a_r \mbox{ fm}$    \\
\hline
Cornell - I                  & (\ref{e:cornell-a})      & $0.52$     &  $2.34$                        &  $5.17$                       &  $-0.50805$            &  -                                   &          -           &       -        &                     -                       &        -                       \\
Cornell - II                 & (\ref{e:cornell-b})      & $0.52$     &  $2.34$                        &  $5.17$                       &  $-0.50805$            &  -                                   &         $1$       &  $0.923$ &                      -                      &        -                      \\
Screened Cornell -I  & (\ref{e:screened1a})  & $0.423$   &  $1.858$                      &  $4.6645$                  &  -                              &  $0.71$                         &             -         &        -      &                       -                     &        -                      \\
Screened Cornell -II & (\ref{e:screened1b})  & $0.423$   &  $1.858$                      &  $4.6645$                  &  -                              &  $0.71$                         &    $0.810$      &      $1$   &                        -                    &        -                      \\
Richardson              & (\ref{e:richardson})    & -               &  -                                  &  $4.8877$                  &  -                              &  -                                  &            -          &         -     & $0.398$                               &         $0.1782$       \\
\hline\hline
\end{tabular}
\end{table}

\subsection{The p-QCD model}

One last potential model we consider in our study \cite{Gupta:1982kp,Radford:2007vd} is based on a semi-relativistic treatment of perturbative QCD interactions to one loop, which we call 
the \emph{p-QCD model}, where the spin-dependent potentials, which appear in the perturbative treatment, are somewhat different than those  in the previous models.
The Hamiltonian in this model considers a relativistic expression for the quark  kinetic
energy, $H_0 = \sqrt{\vec{p}^2+m^2}$, unlike the previous models where a purely non-relativistic approximation, $H_0 = m + p^2/2m$, is used. Additionally, the potential is composed of a short distance part obtained from  perturbative QCD to one loop, and a phenomenological  long range confining part.
The complete potential has the form
\begin{eqnarray}
V(r) = - \frac{4\alpha_S}{3r} \left[1-\frac{3\alpha_S}{2\pi}+\frac{\alpha_S}{6\pi}(33-2n_f) \left(\ln \mu r+\gamma_E\right)\right] +
 A \, r + V_S +  V_L .\label{RR}
 \end{eqnarray}
The first term is the spin independent Coulomb-like part, corrected to one loop in QCD. 
The term  $A \, r$, where $A$ is a phenomenological constant, is the spin-independent,  long range confining potential, assumed to be linear in $r$. 
$V_S$ is a short-range part of the potential that appears in the perturbative calculation to one loop, and includes most spin-dependent parts in the form of spin-orbit (LS), tensor (T) and hyperfine (HF) terms, and a short-interaction spin independent term (SI):
%
\begin{eqnarray}
V_{LS}&=&\frac{2\al_S\mathbf{L}\Dot\mathbf{S}}{m^2r^3}\!\left\{1- \frac{\al_S}{6\pi}\left[\frac{11}{3}-(33-2n_f)\left(\ln\mu r+\ga_E-1\right)+12\left(\ln mr+\ga_E-1\right)\right]\right\}\label{potb}, \nonumber\\ [4pt]
V_{T\;}
&=&\frac{4\al_S(3\mathbf{S_1}\Dot\mathbf{\hat{r}}\mathbf{S_2}\Dot\mathbf{\hat{r}}-\mathbf{S_1}\Dot\mathbf{S_2})}
{3m^2r^3}\left\{1+\frac{\al_S}{6\pi}\left[8+(33-2n_f)\left(\ln\mu
r+\ga_E-\frac{4}{3}\right) -18\left(\ln mr+\ga_E-\frac{4}{3}\right)\right]\right\},\label{potc} \nonumber\\
V_{HF}&=&\frac{32\pi\al_S\mathbf{S_1}\Dot\mathbf{S_2}}{9m^2}\left\{\left[1-\frac{\al_S}{12\pi}
(26+9\ln\,2)\right]\de(\mathbf{r})\right. \nonumber\\
& & \left.-\frac{\al_S}{24\pi^2}(33-2n_f)\nabla^2\left[\frac{\ln\,\mu r+\ga_E}{r}\right]+\frac{21\al_S}{16\pi^2}\nabla^2\left[\frac{\ln\, mr+\ga_E}{r}\right]\right\}\label{pota}, \nonumber \\
V_{SI}&=&\frac{4\pi\al_S}{3m^2}\left\{\left[1-\frac{\al_S}{2\pi}(1+\ln2)\right]
\de(\mathbf{r})-\frac{\al_S}{24\pi^2}(33-2n_f)\nabla^2\left[\frac{\ln\,\mu
r+\ga_E}{r}\right]-\frac{7\al_Sm}{6\pi r^2}\right\}. \label{potd}\nonumber
\end{eqnarray}
%
Finally $V_L$ is the relativistic correction arising from the confining potential (thus proportional to $A$), also composed of spin-dependent terms:
\begin{equation}
V_L = -(1-f_V)\frac{A}{2m^2 r}\mathbf{L}\cdot\mathbf{S}
+ f_V\left[\frac{A}{2m^2 r}\left(1+\frac{8}{3}\mathbf{S_1}\cdot \mathbf{S_2}\right)
+ \frac{3A}{2m^2 r}\mathbf{L}\cdot \mathbf{S} + \frac{A}{3m^2 r}(3\mathbf{S_1}\cdot \mathbf{\hat{r}}\mathbf{S_2}\cdot \mathbf{\hat{r}} - \mathbf{S_1}\cdot \mathbf{S_2})
\right].      
\nonumber
\end{equation}

The phenomenological parameter  $f_V$ was introduced by the authors to represent the fraction of vector vs. scalar character of the confining potential, which they fit with the available data.
For bottomonium the parameters $A$, $m$, $\alpha_S$, $\mu$ and $f_V$ were fitted using as input eight masses of the $b\bar{b}$ spectrum. The authors  obtained
 \cite{Radford:2009qi} $A=0.175\mbox{ GeV}^2$, $m=5.33\mbox{ GeV}$, $\alpha_S=0.295$,  
 $\mu=4.82\mbox{ GeV}$ and $f_V=0$. The values of $\alpha_S$ and $m$ are given in the renormalization scheme
 of Ref.~\cite{Gupta:1982im} for the value of $\mu$  given above.

Remarkably enough, $f_V=0$ implies, just like in the other models, that the long distance part of the potential corresponds to purely scalar exchange.

\section{Calculations and Results} \label{results}

\subsection{The spin-average masses of $\chi_b(1P)$, $\chi_b(2P)$ and $\chi_b(3P)$}

 As seen in Eq.\ (\ref{e:splittings}), the masses of the states $\chi_b(nP_0)$, $\chi_b(nP_1)$ and $\chi_b(nP_2)$ differ by small amounts. These differences are determined by spin-dependent interactions. 
 On the other hand, the  barycenter (or spin-averaged) mass of these states is defined as
\begin{eqnarray}\label{bary}
\bar{M} = \frac{5 M(P_2) + 3 M(P_1) + M(P_0)}{9}, 
\end{eqnarray}
and is determined in each model by the leading, spin-independent part of the potential.
Consequently, we expect the spin-independent part of the models we previously described to correctly reproduce the experimental barycenters. We have calculated $\bar{M}$ for the $\chi_b(1P)$, $\chi_b(2P)$ and $\chi_b(3P)$, numerically solving the Schr\"odinger equation \cite{Lucha:1998xc} for the potential models described in the previous section, 
using the parameters listed in Table \ref{t:parameters1}, for the first three models, respectively.
The results we obtain are shown in Table \ref{t:avgmasses}. We include the results obtained in Ref.\  \cite{Radford:2009qi} by the authors of the p-QCD model.
\begin{table}[h]
\caption{The mass barycenters of the states $\chi_{b}(1P)$, $\chi_{b}(2P)$ and $\chi_{b}(3P)$ reproduced by the potential models described in Section \ref{models}, and compared to the corresponding experimental values.
On the right of each reproduced mass is the discrepancy between this value and the corresponding experimental value, $\delta_{nP}\equiv  \bar M(nP)_{(\mbox{model})}-\bar M(nP)_{(\mbox{exp})}$.
The experimental values for $1P$ and $2P$ are from Ref.\ \cite{pdg:2010-11}, while  those for $3P$ are from Ref.\ \cite{Aad:2011ih}. All values are in MeV. } 
\label{t:avgmasses}
\centering
\begin{tabular}{l l r  l r  l r c }
\hline \hline
Model                       & $\bar{M}(1P)$      & $\delta_{1P}$ \ \ \ & $\bar{M}(2P)$       & $\delta_{2P}$ \ \ \ & $\bar{M}(3P)$\ \ \  & $\delta_{3P}$   & $[\mbox{MeV}]$ \\
\hline
Cornell                     & $9958.3$           & $58.4$        \ \ \ & $10312.6$           & $52.5$        \ \ \ & $10595.3$           & $65.3$      &  \\
Screened Cornell            & $9907.9$           & $8.0$         \ \ \ & $10261.2$           & $1.0$         \ \ \ & $10516.4$           & $-13.6$   &  \\ 
Richardson                  & $9895.7$           & $-4.2$        \ \ \ & $10248.8$           & $-11.4$       \ \ \ & $10520.1$           & $-9.9$   &  \\
p-QCD \cite{Radford:2009qi} & $9898.7$           & $-1.2$        \ \ \ & $10261.2$           & $1.0$         \ \ \ & $10543.9$           & $13.9$   &  \\
Experiment                  & $9899.87 \pm 0.27$ &   -           \ \ \ & $10260.20 \pm 0.36$ &  -            \ \ \ & $10530 \pm 10 $ &  -              &  \\
\hline\hline
\end{tabular}
\end{table}
The value of the constant $V_0$ for the Cornell potential [Eq.\ (\ref{e:cornell-a})]  was fitted in order to obtain the experimental value of the $\Upsilon(1S)$.

We should recall that the parameters in the Cornell and Richardson potentials, which are shown in Table \ref{t:parameters1}, were fitted using charmonium states (except for the bottom constituent mass $m_b$), and those of the Screened Cornell model were fixed using both $c\bar c$ and $b\bar b$ masses. 
In contrast, in the p-QCD model the free parameters were fitted using $b\bar b$ states only. 
Consequently, one should naturally expect the latter model to give a closer prediction for the $\chi_b$, in comparison with the other models, where an accurate prediction for $\chi_b$ masses is clearly a more demanding requirement.

The bottom line of Table \ref{t:avgmasses} shows the experimental values of the barycenter masses, in order to compare them with the model results.   
To ease the comparison we also include, for each $\chi_b$, the mass discrepancy $\delta_{nP}$  between the model result and the corresponding experimental value.

As a first observation from Table \ref{t:avgmasses}, one can see that in most cases the model discrepancies are larger than the experimental uncertainties, indicating that in general there is still need for improvement in the models. 

Notwithstanding, the errors are quite small in all models except \emph{Cornell}. This tendency indicates that the inclusion of saturation (i.e. going from Cornell to Screened Cornell) clearly improves the predictions for the barycenter masses, but there is no clear distinction in the reliability between the last three models at this stage.

We must also comment that the discrepancy in the Cornell model can be considerably improved in a very simple way: the offset $V_0$ was fitted using the $\Upsilon(1S)$ mass. If we had fitted $V_0$ using $\chi_b(1P)$ instead, the discrepancy in the barycenter mass of $\chi_b(2P)$ and $\chi_b(3P)$ in Table \ref{t:avgmasses}  would have been $\delta_{2P} = - 6.1$ MeV  and $\delta_{3P} = 6.8$ MeV, instead of $52.5$ and $65.3$ MeV, respectively.  In this sense, the Cornell potential does not fit very well the whole $b\bar b$ spectrum, but could still be good in predicting mass \emph{differences} within $\chi_b(nP)$ states, which are all of the same spin and orbital angular momentum. 

\subsection{The splittings of  $\chi_b(1P_J)$,   $\chi_b(2P_J)$ and $\chi_b(3P_J)$}\label{s:splittings}

To date, the splittings of the $J=0,1,2$ states in the $\chi_b(1P_J)$ and $\chi_b(2P_J)$ have been experimentally measured with reasonable precision, while there is still no data 
for the splitting in the $\chi_b(3P_J)$.  Our main goal here is to predict the latter splitting according to each of the aforementioned models. 

We characterize the splittings of the $J=0, 1, 2$ states within a given $\chi_b(nP)$ by the mass differences
$\Delta M_{21}(nP)$ and $\Delta M_{10}(nP)$, as defined in Eq.\ (\ref{SplitD}).   
An additional parameter which is used in the literature to characterize the splittings is the ratio of the mass differences  within a given $\chi_b(nP)$, defined as
\beqa
R_{\chi}(nP) = \frac{\Delta M_{21}(nP)}{\Delta M_{10}(nP)}.\label{Splitx}
\eeqa

In our calculations, we first reproduce within each model the 
splittings in the  $\chi_b(1P_J)$ and $\chi_b(2P_J)$ and compare these results with their corresponding experimental values, as a way to check the reliability of the models. These results are shown in Tables  \ref{t:1psplit} and \ref{t:2psplit}, respectively. 
In these tables one can make several observations. 

First, from Table \ref{t:1psplit}  for the splitting in  $\chi_b(1P)$, the best results for the  $\Delta M_{ij}$ are obtained with the Richardson and p-QCD models, followed by the Cornell-II model, while the largest discrepancies appear with the Screened Cornell-I model. On the other hand, concerning the splitting ratios $R_\chi$, the pattern of performance is different: the Screened Cornell-I  joins the Richardson and p-QCD models in giving the best results, while the largest discrepancy occurs with the Screened Cornell-II model. 

Alternatively, we can see whether the results for the splitting in $\chi_b(1P)$ improve in going from the Cornell to the Screened Cornell model (i.e. by including saturation at long distances), as it happened in the results for the barycenters. Clearly this is not the case: there is no such an improvement on the splittings: the Screened Cornell models actually perform worse than the simple Cornell models. 

Finally, we can see whether the best fitted vector vs. scalar separation causes an improvement in the results (i.e. going from models of type -I to type -II). Indeed, 
a moderate improvement can be seen in Table \ref{t:1psplit} going from Cornell-I to -II or from Screened Cornell-I to -II, which in any case is expected, because the fitting was done with that purpose. However, the improvement occurs only in the $\Delta M_{ij}$, but not so in the ratio $R_\chi$.

In turn, for the splittings in $\chi_b(2P)$, the best results for $\Delta M_{ij}$ are given by the p-QCD model, and Screened Cornell-I and -II models, while the largest discrepancy is found  in the results given by the Cornell-I model.  In contrast, for the splitting ratio $R_\chi$ the best results are given by the Screened Cornell I and p-QCD models, while the largest discrepancy occurs in the Screened Cornell-II model. 

From the above one sees that the results are rather disperse: models that reproduce well some of the splitting features do not do so well on other features. 
The inclusion of saturation in the Cornell model does not seem to make it much better, and the optimal adjustment of the vector vs. scalar separation does not cause a significant improvement either. 

The sole exception, to some degree, is the p-QCD model, which tends to be more often within the best results. However, this tendency was expected, since this model contains a larger number of parameters and they were fitted purely with $b\bar b$. In contrast, all the other models were built using the charmonium spectrum and are now required to fit the $b\bar b$ spectrum as well. 

One clear pattern in the results for $\chi_b(1P)$ and $\chi_b(2P)$ is that all models give values for the ratio $R_\chi$  which are consistently  larger than experiment. This feature could be an indication that the separation of the potentials purely in terms of scalar and vector exchanges may not be sufficient, and an expansion into exchanges of additional spin and parity may be required \cite{Franzini:1992nk}.

Now, concerning the prediction of the models for the so far unknown splitting in the $\chi_b(3P_J)$, Table \ref{t:3psplit} shows the results of our calculations, together with the prediction of the p-QCD model according to its authors \cite{Radford:2012ue}.

Looking at the values for the ratio $R_\chi(3P)$, if we extrapolate from the results on $\chi_b(1P)$ and $\chi_b(2P)$ we may suspect that these are also overestimations, i.e. all models tend to give too large values for $R_\chi$. Within that assumption, we may consider more successful those models that give the smallest prediction for $R_\chi$, which correspond to the p-QCD and the Screened Cornell-I model.

However, if we look at the $\Delta M_{ij}$ results in the 3P, the predictions of the Screened Cornell-I model are much lower than all the others. Then, going back to 1P and 2P, where there is experimental evidence, this model also gives too small splittings in those states. In that sense,  concerning the splittings the reliability of this model is questionable. 

Besides that observation, among the other models there is a tendency in $\Delta M_{10}$ to be near $22\mbox{ MeV}$. For $\Delta M_{21}$ the predictions are more dispersed, but centered around a value of $15\mbox{ MeV}$.
If we consider these two values as the best predictors for the splitting, one then predicts a ratio $R_\chi = 0.68$ for the $\chi_b(3P)$.

Using the definition of the barycenter given in  Eq.\ (\ref{bary}) we can deduce the masses of the states $J=0$, 1 and 2 of the $\chi_(3P_J)$ in terms of the barycenter and splittings:
\begin{eqnarray}
M(P_0) &=&  \bar M - \frac{5\Delta M_{21} + 8 \Delta M_{10}}{9},\nonumber \\
M(P_1) &=&  \bar M - \frac{5\Delta M_{21}-\Delta M_{10}}{9}, \\
M(P_2) &=& \bar M + \frac{4\Delta M_{21}+\Delta M_{10}}{9}.\nonumber
\end{eqnarray}

Using the experimental value and its uncertainty for the barycenter and the estimated splittings, the masses of the three $\chi_b(3P)$ states ($J=0,1,2$) would then be 
$10502\pm 10$ MeV, $10524\pm 10$ MeV and  $10539\pm 10$ MeV, respectively. On the other hand, the threshold for $B^+ B^-$ decay is $10558.5\pm0.3$ MeV. Consequently our estimate of the $J=2$ state is just $20\pm 10$ MeV below the threshold.

\begin{table}[t]
\caption{The mass splittings in the $\chi_b(1P)$ mesons calculated for each of the models described in Section \ref{models}, and their experimental values: 
 the mass differences $\Delta M_{21}$ and $\Delta M_{10}$ are defined as in Eq.\  (\ref{SplitD}), 
and the splitting ratio $R_\chi$ is defined as in Eq.\ (\ref{Splitx}).   } \label{t:1psplit}
\centering
\begin{tabular}{l c c c}
\hline \hline
Model                       & $\Delta M_{21}(1P)\mbox{ [MeV]} \quad $ & $\Delta M_{10}(1P)\mbox{ [MeV]}$ & $R_{\chi}(1P)$\\ \hline
Cornell-I                   & $23.13$                                 & $32.24$                          & $0.717$\\
Cornell-II                  & $20.12$                                 & $29.14$                          & $0.690$\\
Screened Cornell-I          & $16.19$                                 & $25.21$                          & $0.642$\\
Screened Cornell-II         & $20.87$                                 & $28.46$                          & $0.733$\\
Richardson                  & $18.58$                                 & $30.87$                          & $0.602$\\
p-QCD \cite{Radford:2009qi} & $19.3$                                  & $29.94$                          & $0.645$\\ 
Experiment                  & $19.43\pm 0.57$                         & $33.34\pm 0.66$                  & $0.583\pm 0.021$ \\
\hline\hline
\end{tabular}
\end{table}

\begin{table}[t]
\caption{The mass splittings in the $\chi_b(2P)$ mesons   calculated for each of the models described in Section \ref{models}, and their experimental values: 
 the mass differences $\Delta M_{21}$ and $\Delta M_{10}$ are defined as in Eq.\  (\ref{SplitD}), 
and the splitting ratio $R_\chi$ is defined as in Eq.\ (\ref{Splitx}).  } \label{t:2psplit}
\centering
\begin{tabular}{l c c c c}
\hline \hline
Model                       & $\Delta M_{21}(2P)\mbox{ [MeV]} \quad $ & $\Delta M_{10}(2P)\mbox{ [MeV]}$ & $R_{\chi}(2P)$\\ \hline
Cornell-I                   & $19.22$                                 & $26.43$                          & $0.727$ \\
Cornell-II                  & $16.77$                                 & $23.91$                          & $0.701$ \\
Screened Cornell-I          & $12.91$                                 & $19.34$                          & $0.668$ \\
Screened Cornell-II         & $15.91$                                 & $21.45$                          & $0.742$ \\
Richardson                  & $18.27$                                 & $25.36$                          & $0.720$ \\
p-QCD \cite{Radford:2009qi} & $16.40$                                 & $24.30$                          & $0.675$ \\
Experiment                  & $13.5\pm 0.6$                           & $23.5\pm 1.0$                    & $0.574\pm 0.035$ \\
\hline\hline
\end{tabular}
\end{table}

\begin{table}[t]
\caption{The predictions for the mass splittings in the $\chi_b(3P)$ mesons according to each of the models described in Section \ref{models}: 
 the mass differences $\Delta M_{21}$ and $\Delta M_{10}$ are defined as in Eq.\  (\ref{SplitD}), 
and the splitting ratio $R_\chi$ is defined as in Eq.\ (\ref{Splitx}). } \label{t:3psplit}
\centering
\begin{tabular}{l c c c}
\hline \hline
Model                              & $\Delta M_{21}(3P)\mbox{ [MeV]} \quad $ & $\Delta M_{10}(3P)\mbox{ [MeV]}$ & $R_{\chi}(3P)$\\ \hline
Cornell-I                          & $17.7$                                  & $24.0$                           & $0.734$ \\
Cornell-II                         & $15.4$                                  & $21.8$                           & $0.710$ \\
Screened Cornell-I                 & $10.8$                                  & $15.8$                           & $0.683$ \\
Screened Cornell-II                & $12.9$                                  & $17.3$                           & $0.747$ \\
Richardson                         & $17.1$                                  & $22.1$                           & $0.773$ \\
p-QCD \cite{Radford:2012ue} & $14.8$                                  & $22.1$                           & $0.670$ \\
\hline\hline
\end{tabular}
\end{table}

\section{Summary and Conclusions}\label{conclusions}

Recently, the $\chi_b(3P)$ state was observed and its barycenter was determined, while there is still no available data regarding its mass splittings.

Here we have used four well known potential models to predict the so far unresolved splitting in the $\chi_b(3P)$ states. 
The models were described in Section \ref{models} and are called here the \emph{Cornell}, \emph{Screened Cornell}, \emph{Richardson} and \emph{p-QCD} models, respectively. The first three models were originally adjusted to reproduce the charmonium spectrum, and are required to work for bottomonium as well. In contrast, the latter model was fitted using bottomonium states only.

As a first attempt to test the reliability of these models, we use them to reproduce the known barycenter masses of the $\chi_b(1P)$, $\chi_b(2P)$ and $\chi_b(3P)$ states.

The Cornell model, which is the simplest of all of them, shows the largest deviations in its results for the barycenter masses. 
The Screened Cornell model, which is similar to the previous one, but with the additional feature of 
saturation at long distances, shows a clear improvement in its results for the barycenters.
The other two models, Richardson and p-QCD, give similar values.
Up to here, it seems that the latter three models perform similar in terms of reliability. In any case,
one should notice that the discrepancies in these model results are in general larger than their current experimental uncertainties. This is an indication that there is still need for model improvement.

Now, concerning the splittings, we recall that the barycenters of the $\chi_b$ states are determined by the spin-independent part of the potential, while the splittings are determined by the spin-dependent part. The latter, which is a relativistic correction, should be sub-dominant for heavy quarkonia like the $\chi_b$. The spin-dependent part, however, is not unambiguously determined as an expansion of the potential, as it also depends on the Lorentz character (vector vs. scalar) of the effective interaction. 
In the models we study here, we have used a common phenomenological treatment of this feature, which is a separation of the potential into a vector and a scalar exchange, where it is usually assumed that the vector part should dominate the short distance regime, while the scalar part should represent the confining regime at longer distances. 

Accordingly, in the cases of the Cornell and the Screened Cornell models, we considered two variants for the scalar vs. vector separation. Our variant I considers the commonly assumed separation where the vector part corresponds to the short distance Coulombic term and the scalar part corresponds to the long distance linear (or linear-saturated) term. Our variant II uses an optimal scalar-vector separation that minimizes the error in reproducing the known splittings in the $\chi_b(1P_J)$ and $\chi_b(2P_J)$.
Interesting enough, we found that this best fitted separation is very close to the usually assumed separation in variant I, in which the short distance part is purely vector exchange and the confining long distance part is purely scalar. This result also coincides with a similar analysis done in the p-QCD model by its authors, where they found that their best fit corresponds to a long distance part being purely scalar. 

For the Richardson potential, unlike the previous models, there is no obvious separation into scalar and vector parts, so we introduced a purely phenomenological function with one parameter to carry out a short distance vs. long distance separation.
This parameter, also fitted with the $\chi_b(1P)$ and $\chi_b(2P)$ states, is a length scale that separates the two regions; the short distance part is then assumed to be mediated by vector exchange and the long distance part by scalar exchange.

Of all the models studied, those that give more consistent results for the splittings are the p-QCD model and, to some degree, the Richardson model. 
One could have expected the p-QCD model to give the best results because of its more detailed short-distance treatment from perturbative QCD, and because it contains a larger number of parameters, all fitted to the bottomonium spectrum.  

Concerning the comparison between the Cornell and Screened Cornell models, we would have expected an improvement in the reproduction of the splittings in the latter, as was the case for the reproduction of the barycenters. However it was not the case: we found the Screened Cornell model  to give results that are no better than the simpler Cornell model. On the other hand, when comparing the variants I and II of these models, we had expected better results in the variants II, because they were fitted to do so. This was indeed the case in the results for the $\Delta M_{ij}$, but not so for the ratios $R_\chi$. 

The one consistent pattern in all results for the splittings is that all the models give ratios $R_\chi$ in $\chi_b(1P)$ and $\chi_b(2P)$ that are larger than the corresponding experimental values. Concerning the reliability of the models, this tendency may be an indication that the composition of the potential in a scalar and a vector part may not be enough, and exchanges of further spins and parities may be present. From the point of view of predictions, on the other hand, we may then expect that the predictions for the $\chi_b(3P)$ splittings will follow that tendency, i.e. the model with best predictions should give the smallest value for $R_\chi$. This tendency again falls on the p-QCD model. 

Now, looking at the predictions for the $\Delta M_{10}$ and $\Delta M_{21}$ in $\chi_b(3P)$, one can see that the largest deviations around a central tendency are found in the Cornell-I and Screened Cornell-I models. Besides those predictions the tendency in $\Delta M_{10}$ is to be near $22\mbox{ MeV}$, 
and for $\Delta M_{21}$ the predictions, although more dispersed,  are centered around a value of $15\mbox{ MeV}$. With these estimates and the experimental value for the barycenter, the masses of the three $\chi_b(3P)$ states ($J=0,1,2$) would be $10502\pm 10$ MeV, $10524\pm 10$ MeV and  $10539\pm 10$ MeV, respectively (the model uncertainty is not included). This means that the $J=2$ state would be just $20\pm 10$ MeV below the $B^+B^-$ threshold.

Finally, if we consider these values for $\Delta M_{21}$ and $\Delta M_{10}$ as the best predictions for the splitting, one then predicts a ratio $R_\chi = 0.68$ for the $\chi_b(3P)$, which is again quite close to the result of p-QCD.

\section*{Acknowledgments}
We are grateful to Will Brooks, Ryan White and Franz Sch\"{o}berl for useful discussions. This work was supported in part by Conicyt, Chile grant \emph{Institute for Advanced Studies in Science and Technology} ACT-119.

%
%

\end{document}